\documentclass[aps,pre,showpacs,noshowkeys,amsmath,amssymb,amsfonts,superscriptaddress,longbibliography,reprint]{revtex4-1}
\usepackage[english]{babel}

\usepackage{graphicx}
\usepackage{bm}
\usepackage[colorlinks=true,linkcolor=blue,citecolor=blue,urlcolor=blue]{hyperref}
\usepackage[all]{hypcap}
\usepackage[nameinlink,capitalise]{cleveref}
\begin{document}
\title{Role of Substrate Stiffness in Tissue Spreading: Wetting Transition and Tissue Durotaxis}
\author{Ricard Alert}
\email{ricard.alert@princeton.edu}
\affiliation{Departament de F\'{i}sica de la Mat\`{e}ria Condensada, Universitat de Barcelona, 08028 Barcelona, Spain}
\affiliation{Universitat de Barcelona Institute of Complex Systems (UBICS), Universitat de Barcelona, Barcelona, Spain}
\affiliation{Princeton Center for Theoretical Science, Princeton University, Princeton, NJ 08544, USA}
\affiliation{Lewis-Sigler Institute for Integrative Genomics, Princeton University, Princeton, NJ 08544, USA}
\author{Jaume Casademunt}
\email{jaume.casademunt@ub.edu}
\affiliation{Departament de F\'{i}sica de la Mat\`{e}ria Condensada, Universitat de Barcelona, 08028 Barcelona, Spain}
\affiliation{Universitat de Barcelona Institute of Complex Systems (UBICS), Universitat de Barcelona, Barcelona, Spain}
\date{\today}

\begin{abstract}
Living tissues undergo wetting transitions: On a surface, they can either form a droplet-like cell aggregate or spread as a monolayer of migrating cells. Tissue wetting depends not only on the chemical but also on the mechanical properties of the substrate. Here, we study the role of substrate stiffness in tissue spreading, which we describe by means of an active polar fluid model. Taking into account that cells exert larger active traction forces on stiffer substrates, we predict a tissue wetting transition at a critical substrate stiffness that decreases with tissue size. On substrates with a stiffness gradient, we find that the tissue spreads faster on the stiffer side. Further, we show that the tissue can wet the substrate on the stiffer side while dewetting from the softer side. We also show that, by means of viscous forces transmitted across the tissue, the stiffer-side interface can transiently drag the softer-side interface towards increasing stiffness, against its spreading tendency. These two effects result in directed tissue migration up the stiffness gradient. This phenomenon --- tissue durotaxis --- can thus emerge both from dewetting at the soft side and from hydrodynamic interactions between the tissue interfaces. Overall, our work unveils mechanisms whereby substrate stiffness impacts the collective migration and the active wetting properties of living tissues, which are relevant in development, regeneration, and cancer.
\end{abstract}

\maketitle

\section{Introduction}

In embryonic development, wound healing, and tumor progression, epithelial cells migrate collectively in cohesive groups \cite{Friedl2009}. To study the mechanics of collective cell migration, extensive research has focused on the spreading of epithelial tissues in vitro \cite{Hakim2017a}. For example, when a cell aggregate is placed on a surface, it may retain a spheroidal shape or it may spread by extending a cell monolayer \cite{Gonzalez-Rodriguez2012,Beaune2014}. Similar situations are found in vivo, for example during the epiboly process in zebrafish embryogenesis \cite{Morita2017a,Wallmeyer2018}. In analogy with the wetting of a liquid drop, the spreading of cell aggregates was proposed to rely on a competition between cell-cell and cell-substrate interactions \cite{Ryan2001,Douezan2011,Ravasio2015a,Smeets2016}. In experiments, these surface interactions were varied by tuning the expression level of cell-cell adhesion proteins and by modifying the chemical coating of the substrate, respectively. Expectedly, these changes can induce a wetting transition between a three-dimensional cell aggregate and a spreading cell monolayer --- the equivalents of a drop and a precursor film, respectively \cite{Douezan2011}.

However, the wetting behavior of a living tissue is not completely analogous to that of an inert liquid. The ability of cells to polarize and exert traction forces to migrate on a substrate turns a tissue into an active material, fundamentally changing the physics of tissue wetting \cite{Perez-Gonzalez2019}. Moreover, cells sense and respond to the mechanical properties of their environment \cite{Discher2005,Ladoux2012,Gupta2016}. As a consequence, substrate stiffness affects the wetting properties of living tissues. Specifically, cell monolayers dewet from very soft substrates \cite{Douezan2012a} but wet stiffer substrates \cite{Douezan2012c,Guo2006a}. In a reminiscent in vivo situation, the developmental stiffening of a tissue triggers the collective migration of neural crest cells required for embryo morphogenesis \cite{Barriga2018a}.

Besides affecting tissue wetting, substrate elasticity also influences the coordination and guidance of collective cell migration \cite{Angelini2010,Ng2012,Saez2007,Edwards2011,Lange2013,Ladoux2017a,Barriga2018b}. Most strikingly, cell monolayers can migrate towards increasing substrate stiffness \cite{Sunyer2016}, a behavior known as tissue durotaxis.

How do these different collective phenomena emerge from the interplay between the mechanical properties of the substrate, cellular forces, and cell-cell interactions? Here, we address this question theoretically. We generalize an active polar fluid model of tissue spreading to account for the adaptation of cellular traction forces to substrate stiffness. This way, we predict a critical substrate stiffness for tissue wetting as a function of active cellular forces and tissue size. Considering a stiffness gradient, we numerically obtain that the tissue spreads faster on the stiffer region, reproducing experimental observations \cite{Sunyer2016}. Moreover, we also unveil two mechanisms for tissue durotaxis. First, the tissue interface on the stiffer side can wet the substrate while the softer-side interface dewets from it. Second, through the transmission of hydrodynamic forces across the tissue, the stiffer-side interface can drag the softer-side interface towards increasing stiffness, counteracting its local spreading tendency. We analytically predict the presence or absence of this dragging effect as a function of the stiffness gradient, the size of the tissue, and its position on the gradient.

\section{Theoretical model}

\begin{figure}[tb]
\begin{center}
\includegraphics[width=\columnwidth]{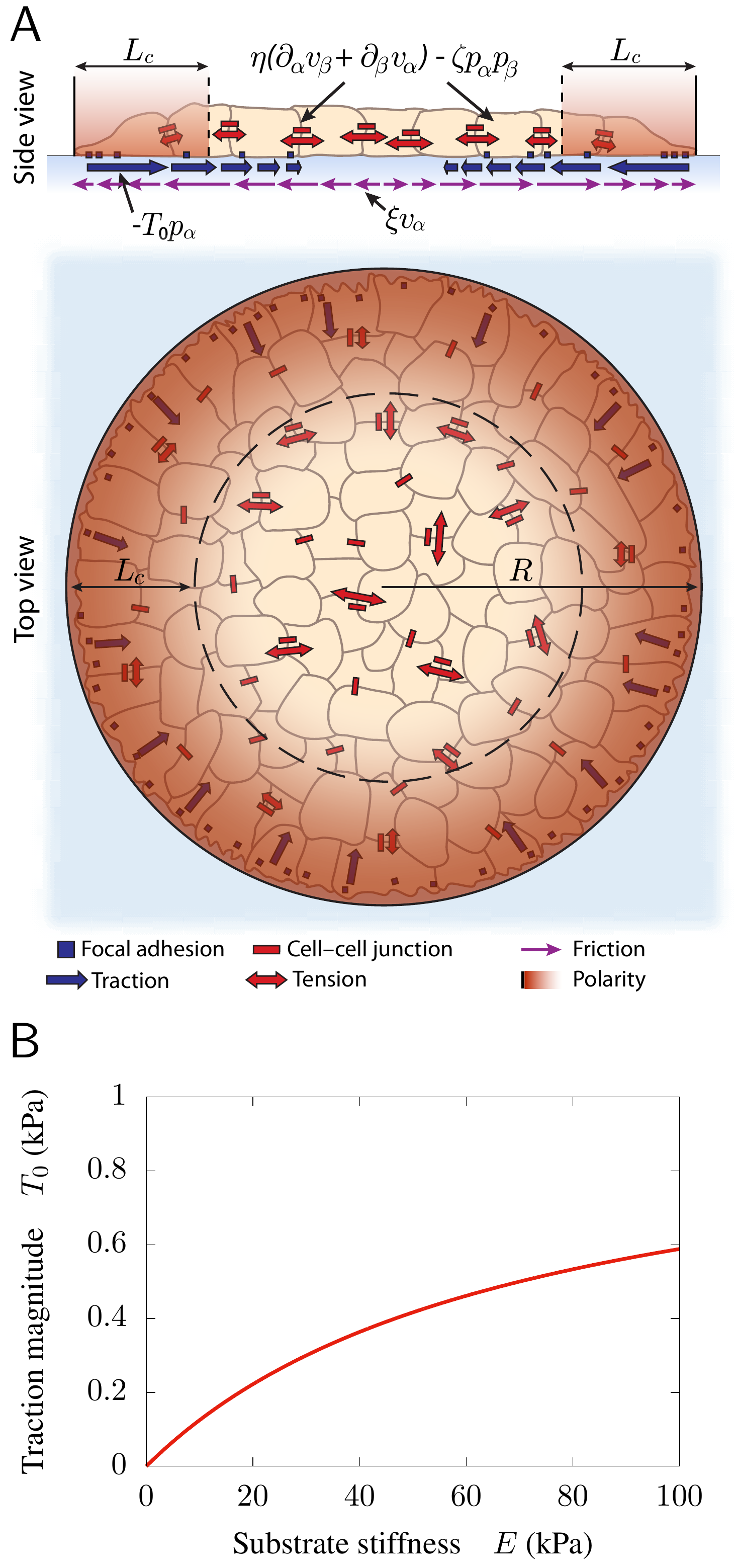}
\caption{Active polar fluid model of tissue wetting. (A) Scheme of the model. From \cite{Perez-Gonzalez2019}. (B) The magnitude of traction forces increases with substrate stiffness, \cref{eq mechanosensing}. Parameter values are in \cref{t parameters}.} \label{fig model}
\end{center}
\end{figure}

We base our analysis on a continuum active polar fluid model for the spreading of an epithelial monolayer, which is thus described in terms of a polarity field $\vec{p}(\vec{r},t)$ and a velocity field $\vec{v}(\vec{r},t)$ in two dimensions \cite{Blanch-Mercader2017,Perez-Gonzalez2019}. We neglect cell proliferation as well as the bulk elasticity of the cell monolayer, which eventually limit the spreading process \cite{Serra-Picamal2012,Basan2013,Recho2016,Yabunaka2017a}. Tissue spreading is driven by the traction forces exerted by cells close to the monolayer edge, which polarize perpendicularly to the edge to migrate towards free space. In contrast, the inner region of the monolayer remains essentially unpolarized, featuring much weaker and transient traction forces \cite{Perez-Gonzalez2019} (\cref{fig model}A). Hence, we take a free energy for the polarity field that favors the unpolarized state $p=0$ in the bulk, with a restoring coefficient $a$, and we impose a normal and maximal polarity as a boundary condition at the monolayer edge. In addition, the polar free energy includes a cost for polarity gradients, with $K$ the Frank constant of nematic elasticity in the one-constant approximation \cite{DeGennes-Prost}. Altogether,
\begin{equation}
F=\int \left[\frac{a}{2}p_\alpha p_\alpha + \frac{K}{2}(\partial_\alpha p_\beta)(\partial_\alpha p_\beta)\right]\,\mathrm{d}^3\vec{r}.
\end{equation}
We assume that the polarity field is set by flow-independent mechanisms, so that it follows a purely relaxational dynamics, and that it equilibrates fast compared to the spreading dynamics \cite{Perez-Gonzalez2019}. Hence, $\delta F/\delta p_\alpha=0$, which yields
\begin{equation} \label{eq polarity-field}
L_c^2 \nabla^2 p_\alpha=p_\alpha,
\end{equation}
where $L_c=\sqrt{K/a}$ is the characteristic length with which the polarity modulus decays from $p=1$ at the monolayer edge to $p=0$ at the center (red shade in \cref{fig model}A).

Then, force balance imposes
\begin{equation} \label{eq force-balance}
\partial_\beta \sigma_{\alpha\beta} + f_\alpha=0,
\end{equation}
where $\sigma_{\alpha\beta}$ is the stress tensor of the monolayer, and $f_\alpha$ is the external traction force density acting on it. We relate these forces to the polarity and velocity fields via the following constitutive equations for a compressible active polar fluid \cite{Perez-Gonzalez2019,Oriola2017}:
\begin{subequations} \label{eq constitutive-equations}
\begin{align}
\sigma_{\alpha\beta} &= \eta\left(\partial_\alpha v_\beta + \partial_\beta v_\alpha\right) - \zeta p_\alpha p_\beta,\\
f_\alpha &= -\xi v_\alpha + \zeta_i p_\alpha.
\end{align}
\end{subequations}
Here, $\eta$ is the monolayer viscosity, and $\xi$ is the cell-substrate friction coefficient. Respectively, $\zeta<0$ is the active stress coefficient accounting for the contractility of polarized cells, and $\zeta_i>0$ is the contact active force coefficient accounting for the maximal traction stress exerted by polarized cells on the substrate, $T_0=\zeta_i h$, with $h$ the monolayer height (\cref{fig model}A).

To model cellular response to substrate properties, we assume that the parameters of cell-substrate interactions depend on substrate stiffness. Tissue cells tend to exert larger traction forces on stiffer substrates \cite{Discher2005,Ladoux2012,Gupta2016,Saez2005,Ghibaudo2008,Ladoux2010,Trichet2012,Elosegui-Artola2014,Gupta2015,Li2015b,Saez2010}. Many studies have explained this response by simply assuming that intracellular force is exerted on an in-series connection of two linear elastic media, namely the substrate and the attached cellular structures, with Young's modulus $E$ and $E^*$, respectively \cite{Walcott2010,Zemel2010,Marcq2011,Trichet2012,Sens2013a,Gupta2015}. Following them, we take cell-substrate forces that depend on substrate stiffness as (\cref{fig model}B)
\begin{equation} \label{eq mechanosensing}
T_0(E)=T_0^\infty \frac{E}{E+E^*}, \qquad \xi(E)=\xi_\infty \frac{E}{E+E^*},
\end{equation}
where $T_0^\infty$ and $\xi_\infty$ are the maximal active traction stress and friction coefficient on an infinitely stiff substrate, respectively.

This minimal approach neglects some aspects of the cellular response to the mechanical properties of the substrate. First, we focus on purely elastic substrates; we do not consider substrate viscoelasticity, which also affects cellular and tissue forces \cite{Murrell2011,Chaudhuri2015,Imai2015,Zheng2017,Bennett2018,Charrier2018}. Second, we do not explicitly account for the substrate deformation field, which could mediate long-range elastic interactions between cells \cite{Angelini2010,Reinhart-King2008}. However, a dependence very similar to \cref{eq mechanosensing} and \cref{fig model}B was obtained when accounting for the non-local substrate elasticity \cite{Banerjee2012}. Finally, cell-cell and cell-substrate adhesions are coupled through the actin cytoskeleton, and hence they may exhibit mechanical crosstalk. However, the influence of substrate stiffness on cell-cell interactions \cite{Guo2006a,Ng2012,Lembong2017,Ladoux2017a,Perez-Gonzalez2019} remains poorly understood. Thus, for the sake of simplicity, we assume a stiff\-ness-in\-de\-pen\-dent intercellular contractility $\zeta$.

\begin{table}[tb]
\begin{center}
\begin{tabular}{clc}
Symbol&Description&Estimate\\\hline
$h$&monolayer height&$5$ $\mu$m \cite{Perez-Gonzalez2019}\\
$L_c$&nematic length&$25$ $\mu$m \cite{Blanch-Mercader2017,Perez-Gonzalez2019}\\
$T_0^\infty$&maximal traction&$1$ kPa \cite{Blanch-Mercader2017,Perez-Gonzalez2019}\\
$-\zeta$&intercellular contractility&$20$ kPa \cite{Perez-Gonzalez2019}\\
$\xi_\infty$&maximal friction coefficient&$2$ kPa$\cdot$s/$\mu$m$^2$ \cite{Cochet-Escartin2014}\\
$\eta$&monolayer viscosity&$80$ MPa$\cdot$s \cite{Blanch-Mercader2017,Perez-Gonzalez2019}\\
$\lambda_\infty$&minimal hydrodynamic screening length&$200$ $\mu$m ($\sqrt{\eta/\xi_\infty}$)\\
$E^*$&characteristic cellular stiffness&$70$ kPa \cite{Douezan2012c}\\
$E'$&stiffness gradient&$50$ kPa/mm \cite{Sunyer2016}\\
$E_0$&stiffness offset (soft edge)&$20$ kPa \cite{Sunyer2016}
\end{tabular}
\end{center}
\caption{Estimates of model parameters.} \label{t parameters}
\end{table}

\section{Results and discussion}

\subsection{Wetting transition}

\begin{figure}[tb]
\begin{center}
\includegraphics[width=\columnwidth]{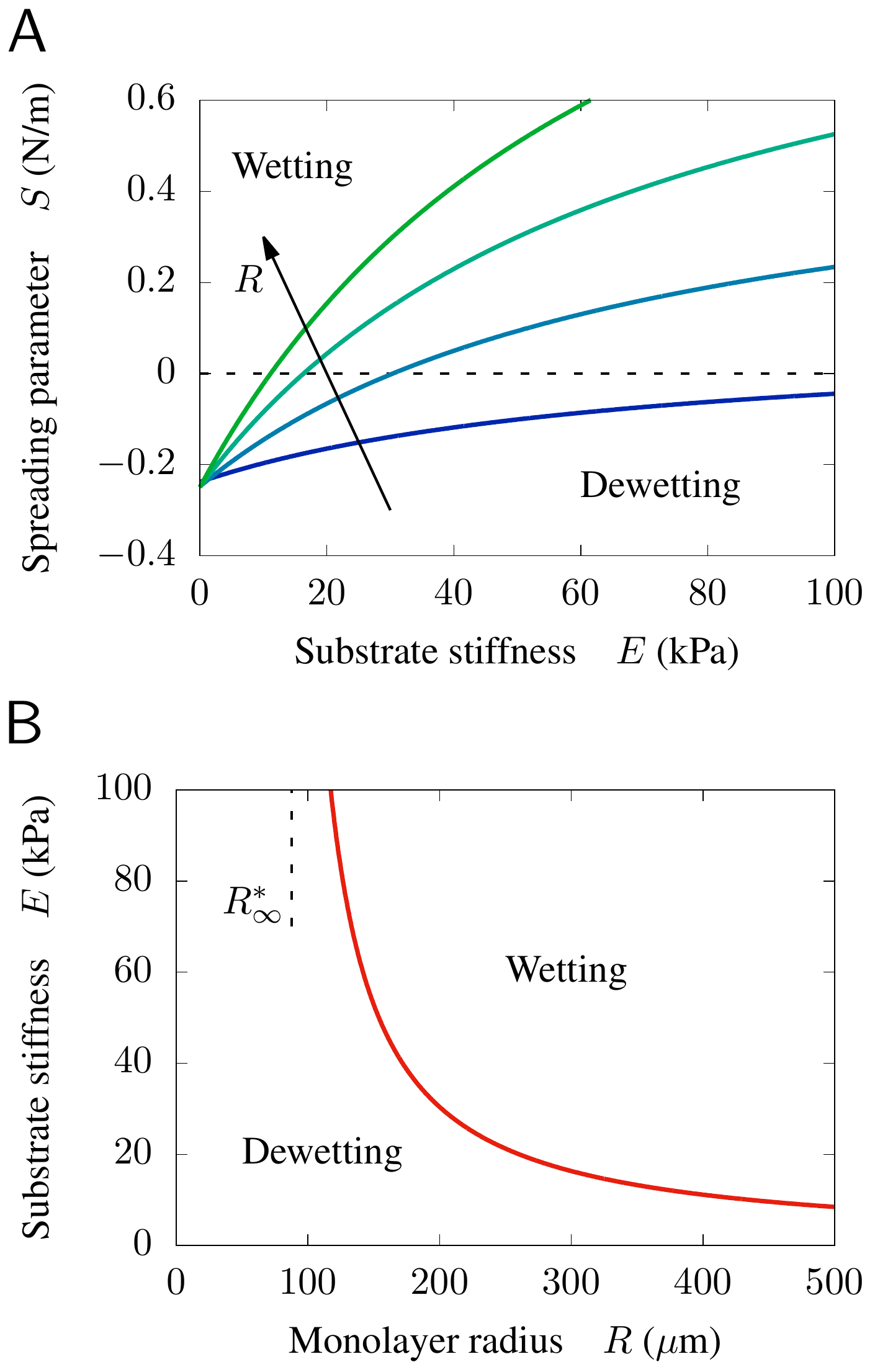}
\caption{Tissue wetting transition induced by substrate stiffness. (A) The spreading parameter of a cell monolayer increases with substrate stiffness and with tissue size, \cref{eq spreading-parameter}. The point at which $S=0$ indicates the critical stiffness for tissue wetting at different monolayer radii $R=100,200,300,400$ $\mu$m (blue to green). (B) The critical stiffness for the wetting transition decreases with monolayer radius, \cref{eq critical-stiffness}. Alternatively, the critical radius for tissue wetting decreases with substrate stiffness, asymptotically tending to $R^*_\infty\approx (-\zeta h/T_0^\infty + 3L_c)/2$. Parameter values are in \cref{t parameters}.} \label{fig wetting}
\end{center}
\end{figure}

In this section, we study the effect of substrate stiffness on the tissue wetting transition. We consider a circular cell monolayer spreading radially (\cref{fig model}A), such as those extending from spheroidal cell aggregates \cite{Beaune2014}. In addition to a maximal normal polarity at the edge, $\vec{p} (R)=\hat{r}$, we impose a stress-free boundary condition, $\sigma_{rr}(R)=0$, with $R$ the monolayer radius. Then, neglecting cell-substrate viscous friction ($\xi\rightarrow 0$), the model \crefrange{eq polarity-field}{eq mechanosensing} can be solved analytically \cite{Perez-Gonzalez2019}. Thus, we obtain the spreading velocity $V=v_r(R)=dR/dt$, and hence the spreading parameter \cite{Beaune2014} $S=\eta V$. For $L_c\ll R$, which is the case in most experiments, it reads
\begin{equation} \label{eq spreading-parameter}
S\approx \frac{T_0^\infty L_c}{h}\frac{E}{E+E^*}\left(R - \frac{3}{2}L_c\right) + \frac{\zeta L_c}{2}.
\end{equation}
The spreading parameter increases with substrate stiffness (\cref{fig wetting}A). Therefore, the competition between intercellular contractility $\zeta<0$ and stiffness-dependent traction forces $T_0(E)>0$ entails the existence of a critical substrate stiffness
\begin{equation} \label{eq critical-stiffness}
E_c\approx E^*\frac{R_\infty^* - 3L_c/2}{R-R_\infty^*}
\end{equation}
above which the tissue spreads ($S>0$, wetting) and below which it retracts ($S<0$, dewetting). The spreading parameter also increases with tissue size (\cref{fig wetting}A), and hence the critical stiffness decreases with monolayer radius (\cref{fig wetting}B). In \cref{eq critical-stiffness}, $R^*_\infty\approx (-\zeta h/T_0^\infty + 3L_c)/2$ is the critical radius for tissue wetting unveiled in our previous work \cite{Perez-Gonzalez2019}, which emerges from the competition between bulk and contact active forces. Here, the critical radius decreases with substrate stiffness,
\begin{equation}
R^*\approx R^*_\infty + \frac{E^*}{E}\left(R^*_\infty - \frac{3}{2}L_c\right),
\end{equation}
asymptotically tending to $R^*_\infty$ for large stiffness (\cref{fig wetting}B).

In the following, we compare our prediction (\cref{eq critical-stiffness}) to published experimental results. For example, for cell monolayers of $R=100$ $\mu$m, and the values of traction and contractility in \cref{t parameters}, P\'{e}rez-Gonz\'{a}lez et al. observed wetting only on substrates of $E=30$ kPa \cite{Perez-Gonzalez2019}, suggesting that $E_c\sim 30$ kPa. Thus, comparing to \cref{eq critical-stiffness}, we estimate that $E^*\sim 30$ kPa for the cell monolayers of Ref. \cite{Perez-Gonzalez2019}. In another study, Douezan et al. reported $E^*\sim 70$ kPa and $E_c\sim 8$ kPa for cell aggregates of radius $R\sim 50$ $\mu$m \cite{Douezan2012c}. Introducing these values into \cref{eq critical-stiffness}, we infer $R_\infty^*\sim 5$ $\mu$m. Taking $h\sim 5$ $\mu$m, this value suggests that the intercellular contractility $-\zeta$ and the maximal traction stress $T_0^\infty$ are of the same order for the cell aggregates of Ref. \cite{Douezan2012c}. In conclusion, our result explains the experimental observation of a wetting transition induced by substrate stiffness \cite{Douezan2012c,Douezan2012a,Perez-Gonzalez2019}, and it predicts that the critical stiffness depends on tissue size. Testing this prediction requires new experiments that, in addition to varying substrate stiffness, would systematically vary tissue size.

\begin{figure*}[tb]
\begin{center}
\includegraphics[width=\textwidth]{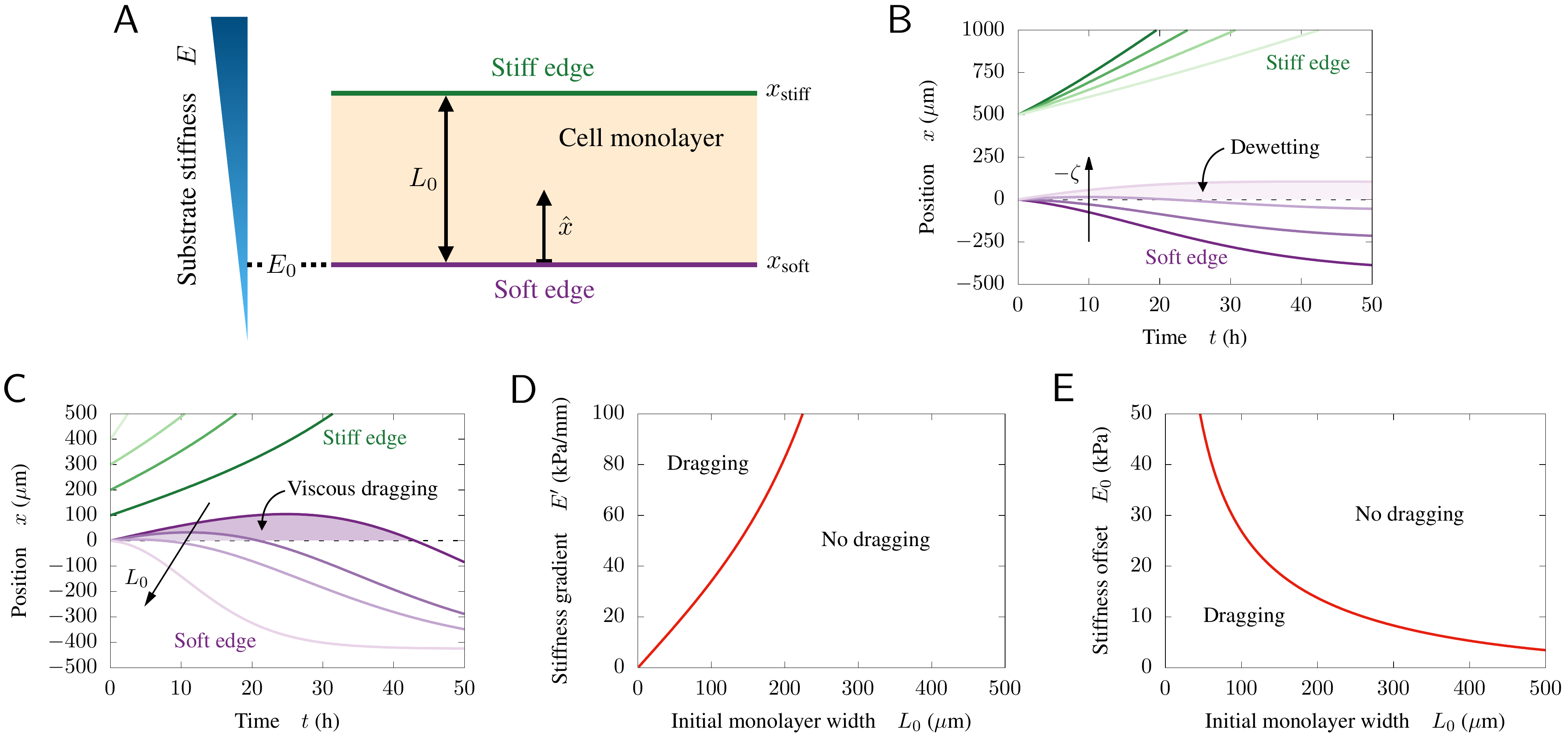}
\caption{Tissue durotaxis. (A) Sketch of a cell monolayer on a substrate with a stiffness gradient. (B) Tissue spreading at increasing intercellular contractility $-\zeta=0,5,10,15$ kPa (darker to lighter). For small contractility, the tissue spreads faster on the stiffer side. For larger contractility, the soft edge retracts (dewetting) while the stiff edge advances (wetting). For this plot, $L_0=500$ $\mu$m $>\lambda_\infty=200$ $\mu$m, so that both tissue interfaces move quite independently. (C) Tissue spreading in the absence of intercellular contractility ($\zeta=0$) at increasing initial monolayer width $L_0=100,200,300,400$ $\mu$m. For narrow monolayers, the stiff edge transiently drags the soft edge towards increasing stiffness by transmission of viscous stress. For wide monolayers, this effect is lost. (D-E) Diagrams of interface dragging. Dragging takes place for sufficiently narrow monolayers, on sufficiently steep stiffness gradients (D), and on sufficiently soft regions of the substrate (E). Except for $\zeta$, parameter values are in \cref{t parameters}.} \label{fig durotaxis}
\end{center}
\end{figure*}

\subsection{Tissue durotaxis}

Motivated by recent experiments \cite{Sunyer2016}, we consider a rectangular cell monolayer spreading on a substrate with a linear stiffness gradient
\begin{equation} \label{eq gradient}
E(x)=E_0+E'x.
\end{equation}
We use the terms ``stiff edge'' and ``soft edge'' to denote the tissue interfaces at the stiffer and softer sides of the substrate, respectively (\cref{fig durotaxis}A). As previously, we impose a normal maximal polarity at the edges, $\vec{p}(x=x_{\text{stiff}},x_{\text{soft}})=\pm\hat{x}$, and stress-free boundary conditions, $\sigma_{xx}(x=x_{\text{stiff}},x_{\text{soft}})=0$. To simulate tissue spreading, we analytically obtain the polarity profile from \cref{eq polarity-field} and we numerically solve the force balance equation, \cref{eq force-balance,eq constitutive-equations} with \cref{eq mechanosensing,eq gradient}, to obtain the velocity profile $v_x(x)$ at each time step. We use a finite differences scheme on a grid of $n=1000$ points and a time step $\Delta t=10$ s. Then, we evolve the interface positions according to $\mathrm{d}x_{\text{stiff}}/\mathrm{d}t=v_x(x_{\text{stiff}})$ and $\mathrm{d}x_{\text{soft}}/\mathrm{d}t=v_x(x_{\text{soft}})$, from initial conditions $x_{\text{stiff}}(0)=L_0$ and $x_{\text{soft}}(0)=0$, respectively.

\subsubsection{Different wetting states at each monolayer edge}


For small intercellular contractility $-\zeta$, active traction forces drive tissue spreading, which is faster on the stiffer side. The soft edge moves towards decreasing stiffness; therefore, its active traction progressively decreases. As a consequence, the soft edge slows down, tending to stop at the substrate position for which contractile tension balances traction forces (\cref{fig durotaxis}B, darker lines).

A larger contractility, however, may overcome the active traction force at the soft edge but not at the stiff edge. In this case, the stiff edge advances but the soft edge retracts (\cref{fig durotaxis}B, lightest lines). Thus, by simultaneously wetting on the stiff side and dewetting from the soft side, the tissue migrates directionally up the stiffness gradient. Finally, a sufficiently large contractility would induce the retraction of both edges, thus causing monolayer dewetting. Overall, these results show that tissue durotaxis can emerge from the different wetting dynamics of the soft and stiff monolayer edges.

\subsubsection{Interface dragging by hydrodynamic force transmission}

The previous durotactic mechanism is at work regardless of tissue size. For monolayers wider than the hydrodynamic screening length $\lambda=\sqrt{\eta/\xi}$, viscous stresses do not fully transmit throughout the tissue because cell-substrate friction screens hydrodynamic interactions at distances larger than $\lambda$ \cite{Blanch-Mercader2017}. Therefore, in the limit $L_0\gg \lambda$, the soft and stiff edges move entirely independently. In contrast, for narrower monolayers, $L_0<\lambda$, both interfaces are strongly coupled.

In the following, we focus on hydrodynamic interactions between the tissue interfaces. To avoid the wetting effects explained above, we set $\zeta=0$. In narrow monolayers, the active tractions on each edge generate flows that span the entire tissue, thus affecting the motion of the other edge. The viscous force transmitted this way can overcome the active traction at the soft edge, outcompeting its spreading tendency and thereby dragging it towards increasing stiffness. This dragging effect is transient; the interfaces eventually decouple when the monolayer becomes too wide to sustain edge-to-edge force transmission (\cref{fig durotaxis}C, darker lines). Accordingly, monolayers that are initially too wide do not experience dragging at all (\cref{fig durotaxis}C, lighter lines). In conclusion, tissue durotaxis can result from interface dragging due to the transmission of viscous forces between the two monolayer edges.

To derive the conditions for interface dragging, we consider the limit in which the width of the polarized boundary layer of cells is much smaller than the total tissue width, $L_c\ll L_0$, which is generally the case in experiments \cite{Blanch-Mercader2017,Perez-Gonzalez2019,Sunyer2016}. In this limit, the active forces are accumulated at the edges, and therefore enter as boundary conditions:
\begin{equation}
\sigma_{xx}(x_{\text{stiff}})=\frac{T_{\text{stiff}}L_c}{h},\qquad \sigma_{xx}(x_{\text{soft}})=\frac{T_{\text{soft}}L_c}{h}.
\end{equation}
Here, $T_{\text{stiff}}=T_0(E(x_{\text{stiff}}))$ and $T_{\text{soft}}=T_0(E(x_{\text{soft}}))$, with $E(x)$ given in \cref{eq gradient} and $T_0(E)$ given in \cref{eq mechanosensing}. Then, force balance reads
\begin{equation}
\frac{\mathrm{d}\sigma_{xx}}{\mathrm{d}x}=\xi v_x,\qquad \sigma_{xx}=2\eta \frac{\mathrm{d}v_x}{\mathrm{d}x},
\end{equation}
which can be solved analytically. From the solution, we determine that the initial velocity of the soft edge is positive (dragging) whenever the following condition is fulfilled:
\begin{equation} \label{eq dragging}
\frac{T^0_{\text{stiff}}-T^0_{\text{soft}}}{T^0_{\text{stiff}}+T^0_{\text{soft}}} > \tanh^2\left(\frac{L_0}{2\sqrt{2}\,\lambda}\right).
\end{equation}
Here, $T^0_{\text{stiff}}=T_0(E(L_0))$ and $T^0_{\text{soft}}=T_0(E(0))$. Thus, in combination with \cref{eq mechanosensing,eq gradient}, \cref{eq dragging} specifies the condition for interface dragging in terms of the stiffness gradient $E'$, the stiffness offset $E_0$, and the initial monolayer width $L_0$. Interface dragging is predicted for monolayers narrow enough to sustain edge-to-edge force transmission. In addition, the monolayer must be on a sufficiently steep stiffness gradient, and on a sufficiently soft region of the substrate to have the required difference in active traction between both edges (\cref{fig durotaxis}D-E).

\subsubsection{Discussion}

In recent experiments, Sunyer et al. observed asymmetric tissue spreading on a gradient of substrate stiffness \cite{Sunyer2016}. Our model reproduces this observation, as we illustrate in \cref{fig durotaxis}B (darker lines). Moreover, on a very soft region of a substrate with a very large stiffness gradient, Sunyer et al. also observed one example of directed tissue migration up the gradient \cite{Sunyer2016}, consistent with our predictions. Here, we have unveiled two mechanisms for such a directed migration, which can be distinguished by their collective character. First, tissue durotaxis based on simultaneous wetting and dewetting (\cref{fig durotaxis}B, lightest lines) needs not be collective in nature; force transmission between edges can take place but is not required. In contrast, durotaxis based on interface dragging (\cref{fig durotaxis}C, darker lines) is a collective effect since it relies on edge-to-edge force transmission. In the experiments by Sunyer et al., durotaxis emerges from intercellular force transmission \cite{Sunyer2016}, so that both mechanisms might be at play. Therefore, further experiments are required to disentangle the contributions of these two durotaxis mechanisms.

Sunyer et al. measured traction forces of equal magnitude on the stiff and soft edges. Hence, their model assumed that cells exert equal active forces on both edges. They then explained durotaxis based on the different substrate deformations caused by these equal forces \cite{Sunyer2016}. However, for almost static cell monolayers, other studies measured larger tractions on stiffer substrates \cite{Saez2010}. This suggests that the active, static contribution of the traction forces increases with substrate stiffness. Hence, here we assumed that cells exert larger active forces on the stiff edge. Thus, in our model, durotaxis is driven by differences in active traction between the tissue edges. Equal total tractions might then result from an additional regulation of cellular forces by stiffness gradients, as well as from the addition of active tractions and viscous friction forces on the substrate, both of which are maximal at the tissue edges and increase with substrate stiffness. Future work could address this point quantitatively by fitting the predictions of our model to experimental data.


Finally, previous mechanical models of tissue durotaxis treated the cell monolayer as an elastic medium, thus imposing full transmission of force through the tissue \cite{Banerjee2011c,Sunyer2016,Gonzalez-Valverde2018,Escribano2018a}. Instead, given that tissue spreading occurs over time scales of several hours \cite{Blanch-Mercader2017}, at which the tissue should have a fluid behavior \cite{Gonzalez-Rodriguez2012,Wyatt2016,Khalilgharibi2016}, we model the cell monolayer as a viscous medium. Thus, our model can suitably account for wetting effects and hydrodynamic interactions in tissue spreading. In particular, we show that these two intrinsic features of fluid media can naturally give rise to durotaxis, which needs not rely on long-range elastic interactions across the tissue. Moreover, we show that the screening of hydrodynamic interactions at long distances gives rise to size-dependent effects such as viscous dragging. Looking for these effects in future experiments may help discriminate between the elastic and viscous behavior of spreading tissues.

\section{Conclusions}

We have studied how substrate stiffness affects the spreading of epithelial tissues. We have extended an active polar fluid model for tissue spreading to incorporate the dependence of cellular traction forces on substrate stiffness. This way, we have shown how substrate stiffness induces a wetting transition between a droplet-like cell aggregate and a spreading monolayer, explaining experimental observations \cite{Douezan2012c,Perez-Gonzalez2019}. We have predicted that the critical stiffness for tissue wetting decreases with tissue size. Further experiments are required to test this prediction. Moreover, we have also explained how gradients of substrate stiffness may give rise to collective cell migration towards increasing stiffness, a behavior known as tissue durotaxis. We have detailed two mechanisms for tissue durotaxis, one based on different wetting states and the other on hydrodynamic interactions between the tissue interfaces. Both mechanisms can coexist, but they can be discriminated because the latter is lost for sufficiently wide cell monolayers. Thus, further systematic experiments are required to assess the relative contributions of the two durotactic mechanisms that we have unveiled. Overall, our results show how the adaptation of cellular traction forces to substrate stiffness impacts the collective migration and the active wetting properties of epithelial tissues.


\begin{acknowledgements}
We thank Carlos P\'{e}rez-Gonz\'{a}lez for discussions and for designing Fig. 1A. We thank Carles Blanch-Mercader, Raimon Sunyer, Xavier Trepat, and the members of Trepat's lab for discussions. R.A. acknowledges support from Fundaci\'{o} ``La Caixa'' and from the Human Frontiers of Science Program (LT000475/2018-C). R.A. thanks Jacques Prost and acknowledges EMBO (Short Term Fellowship ASTF 365-2015), The Company of Biologists (Development Travelling Fellowship DEVTF-151206), and Fundaci\'{o} Universit\`{a}ria Agust\'{i} Pedro i Pons for supporting visits to Institut Curie. R.A. and J.C. acknowledge the MINECO under project FIS2016-78507-C2-2-P and Generalitat de Catalunya under project 2014-SGR-878.
\end{acknowledgements}

\bibliography{Tissues}

\end{document}